# Widespread Excess Ice in Arcadia Planitia, Mars

Ali M. Bramson[1], Shane Byrne[1], Nathaniel E. Putzig[2], Sarah Sutton[1], Jeffrey J. Plaut[3], T. Charles Brothers[4] and John W. Holt[4]

**Corresponding author:** A. M. Bramson, Lunar and Planetary Laboratory, University of Arizona, Kuiper Space Science Building, 1629 E. University Blvd. Tucson, AZ, 85721, USA. (bramson@lpl.arizona.edu)

**Affiliations:**
[1]Lunar and Planetary Laboratory, University of Arizona, Tucson, Arizona, USA.
[2]Southwest Research Institute, Boulder, Colorado, USA.
[3]Jet Propulsion Laboratory, Pasadena, California, USA.
[4]Institute for Geophysics, University of Texas at Austin, Austin, Texas, USA.

**Key points:**
- Terraced craters: abundant in Arcadia Planitia, indicate subsurface layering
- A widespread subsurface interface is also detected by SHARAD
- Combining data sets yields dielectric constants consistent with decameters of excess water ice

**Abstract:**
The distribution of subsurface water ice on Mars is a key constraint on past climate, while the volumetric concentration of buried ice (pore-filling versus excess) provides information about the process that led to its deposition. We investigate the subsurface of Arcadia Planitia by measuring the depth of terraces in simple impact craters and mapping a widespread subsurface reflection in radar sounding data. Assuming that the contrast in material strengths responsible for the terracing is the same dielectric interface that causes the radar reflection, we can combine these data to estimate the dielectric constant of the overlying material. We compare these results to a three-component dielectric mixing model to constrain composition. Our results indicate a widespread, decameters-thick layer that is excess water ice ~$10^4$ km$^3$ in volume. The accumulation and long-term preservation of this ice is a challenge for current Martian climate models.

# 1. Introduction

Information on past Martian climates is contained within the distribution and volume fraction of water ice. The conventional picture of Amazonian mid-latitude ice is that it is young and pore-filling ground ice, which responds quickly to varying orbital and climatic conditions via atmospheric exchange [*Mellon and Jakosky*, 1993; *Schorghofer and Aharonson*, 2005; *Hudson et al.*, 2007; *Bryson et al.*, 2008]. Throughout the last 3 Gyr of Mars' history, these orbital variations dictate where ice is stable – with ice locked up at the poles during low obliquities and mobilized to lower latitudes during higher obliquities [*Head et al.*, 2003].

Models of the distribution of ice stability with the current Martian climate (and today's lower-than-average obliquity) suggest that water ice should be stable poleward of the mid-latitudes (~40-50°) within a meter of the surface, and stable closer to the surface near the poles [*Mellon et al.*, 2004; *Schorghofer*, 2007]. These predictions are supported by inferences of water-equivalent hydrogen content in the upper meter of the surface using the Gamma Ray Spectrometer and Neutron Spectrometer onboard Mars Odyssey [*Boynton et al.*, 2002; *Feldman et al.*, 2002; *Mitrofanov et al.*, 2002].

The current consensus is that ice in the northern mid-latitudes should be "pore-filling", accumulating in and sublimating from shallow pore spaces in the regolith in equilibrium with atmospheric water vapor. However, multiple observations suggest relatively pure excess ice, with higher ice abundances than possible within the porosity of dry regolith. Evidence includes ice-exposing impacts [*Byrne et al.*, 2009; *Dundas et al.*, 2014], radar-detection of massive ice in debris-covered glaciers [*Holt et al.*, 2008; *Plaut et al.*, 2009a], analysis of the latitude-dependence of pedestal craters [*Kadish et al.*, 2009] and thermokarstic expansion of secondary craters [*Viola et al.*, 2015]. The Phoenix lander (at 69° N) excavated both pore-filling and excess ice in the upper centimeters of the surface [*Mellon et al.*, 2009; *Smith et al.*, 2009].

Neither the source nor the emplacement timing of excess ice in the mid-latitudes are well understood. A recent study by *Viola et al.* [2015] of superposed secondary craters in Arcadia Planitia suggests excess ice in this location that is 10s of Myr old. This age is inconsistent with ice stability models, which predict that any ice should periodically sublimate away with obliquity cycles that vary on much shorter timescales. Proposed formation mechanisms for excess ice include burial and preservation of snowpack or glaciers from past orbital periods [*Head et al.*, 2003; *Levrard et al.*, 2004; *Schorghofer*, 2007; *Holt et al.*, 2008; *Schorghofer and Forget*, 2012], development of ice lenses in the subsurface [*Sizemore et al.*, 2015], and thermal contraction of ice in pore spaces to allow increased water vapor diffusion [*Fisher*, 2005].

Understanding ice-emplacement mechanisms and timing is important for understanding Martian climate history. However, to do so one must first constrain the current distribution and concentration of subsurface ice on Mars. Craters provide a way to probe subsurface structure, and concentric terracing within the walls of simple craters (Figure 1) is often indicative of layering within the target material [*Quaide and Oberbeck*, 1968; *Ormö et al.*, 2013]. Here, we use images and digital terrain models (DTMs) from the Mars Reconnaissance Orbiter (MRO) High Resolution Imaging Science Experiment (HiRISE) [*McEwen et al.*, 2010] to determine the morphology and geometry of terraced craters within Arcadia Planitia, and we combine these data with radar sounding from the MRO Shallow Radar (SHARAD) instrument (Figure 1c) to constrain the dielectric properties (and thus composition) of the materials overlying a widespread, radar-detected interface.

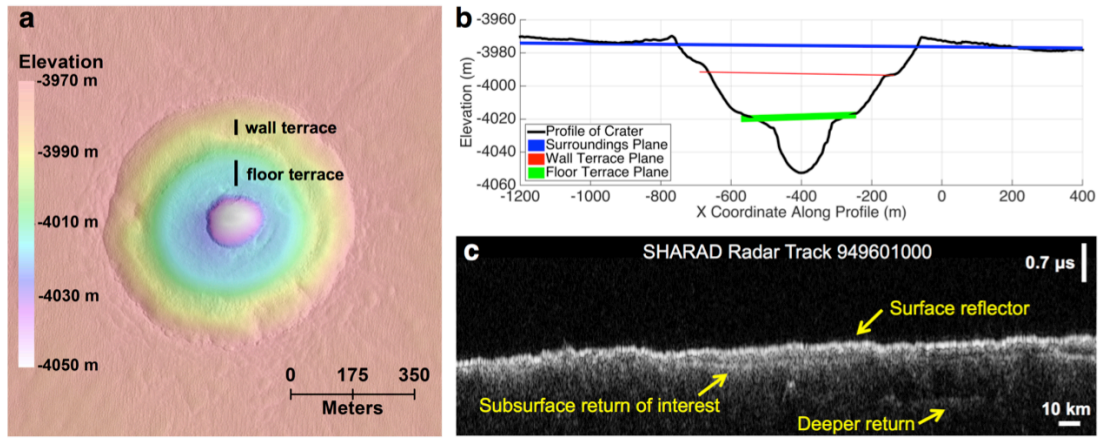

**Figure 1. (a) DTM for a doubly-terraced crater (informally known as Badger Crater) made with HiRISE images ESP_018522_2270 & ESP_019010_2270. (b) Profile across the crater with planes fit to elevations sampled across the surroundings and each terrace. (c) SHARAD radargram that passes within 8.2 km of Badger crater. Vertical scale given in one-way travel time.**

## 2. Methods

We mapped the locations of 187 terraced craters within Arcadia Planitia (Figure 2) using 281 Context Camera (CTX) [*Malin et al.*, 2007] images within the region between 180°-225°E and 38-50°N, chosen on the basis of the boundary within which *Plaut et al.* [2009b] initially found a widespread SHARAD reflector indicative of layering within the subsurface. We projected CTX images using the United States Geological Survey's ISIS (Integrated Software for Imagers and Spectrometers) and Map Projection on the Web tools, and imported the projected images at full-resolution into Esri's geospatial processing program ArcMap™.

We termed a crater as terraced if it exhibited a concentric, flat-lying ledge within the crater wall. Craters were given one of three designations based on the number of terraces and the certainty of terracing: one terrace, two terraces, or questionable terracing. We assigned the latter designation if, at CTX resolution (6 m/pixel), the possible terraces were not pronounced, the concentric structure was only seen in part of the crater, or if the morphology was difficult to distinguish from an expanded crater morphology (as defined by *Viola et al.* [2015]). We followed up on 32 of these terraced craters with HiRISE stereo imaging. From these stereo data, we made 11 DTMs (Figure S3 in supporting information) using a combination of ISIS and the stereogrammetric SOCET SET™ (BAE Systems, Inc.) software, following the method described in *Kirk et al.* [2008] and *Sutton et al.* [2015]. From the DTMs, we sampled the elevations of each terrace at multiple azimuths as well as the surrounding terrain at three crater radii from the crater's center (beyond most ejecta). We fit planes to these elevations for both surroundings and terrace(s), calculating the elevation on each plane at the center of the crater. We recorded the depth of a terrace as the difference between the elevation of the surrounding and terrace planes at the crater's center (details of all measurements are in Table S1 in the supporting information).

We also mapped the surface and subsurface radar interfaces across Arcadia Planitia in 299 SHARAD tracks. These radargrams (along-track images of radar return power versus delay time) were processed with the Smithsonian focused processor [*Campbell et al.*, 2011] on the Colorado SHARAD Processing System (CO-SHARPS) using the processing parameters listed in Text S1. With its 20 MHz center frequency and bandwidth of 10 MHz, SHARAD provides a vertical resolution of 15 m in free space, and $15\varepsilon_r^{-1/2}$ m in a medium of a given dielectric constant ($\varepsilon_r$).

Bright reflections from off-nadir surface topography, known as "clutter", may appear at the same delay times as subsurface returns in the radar data. To check for clutter, we compared the radargrams to corresponding clutter simulations produced using the algorithm of *Holt et al.* [2006, 2008] with Mars Orbiter Laser Altimeter (MOLA) data (Figure S1). We mapped non-clutter subsurface reflectors using SeisWare geophysical interpretation software (SeisWare International Inc.), avoiding radar sidelobes using methods described in Text S1.

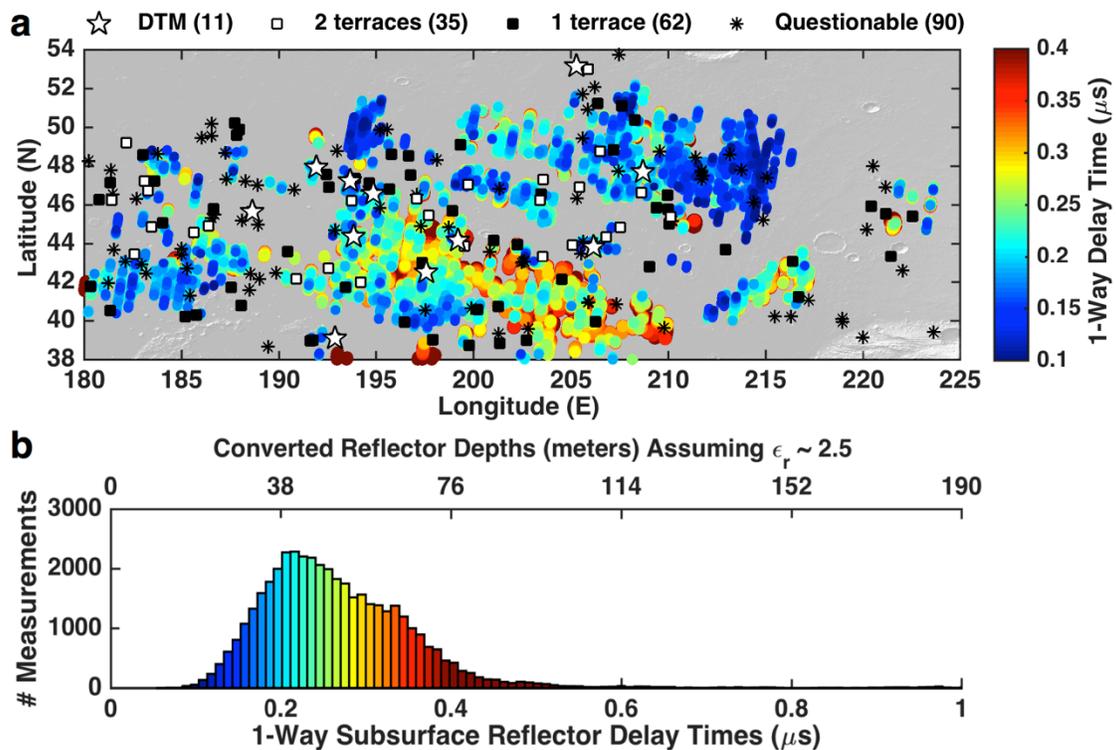

**Figure 2.** (a) Terraced crater locations (B&W) and SHARAD one-way delay times (colors) of mapped subsurface interfaces. (b) One-way delay times in (a).

Assuming that the change in material strength responsible for the terraces corresponds to the dielectric interface that causes the subsurface SHARAD reflection, we can use the terraced crater geometry to independently measure the depth to the subsurface reflector. For each crater, we calculated an inverse-distance weighted mean for all one-way delay times (half of the measured delay times) between the surface and subsurface reflectors within 10 km of the crater. Comparing the depth of the terraces ($\Delta x$) to the time delay between surface and subsurface radar reflections ($\Delta t$), we determined wave velocity (v) and thus the dielectric constant (relative dielectric permittivity) of the overlying material near the craters: $\varepsilon_r = (c\, \Delta t/\Delta x)^2 = (c/v)^2$, where c is the speed of light in vacuum. Text S2 contains our method of estimating errors in $\varepsilon_r$.

The dielectric constant $\varepsilon_r$ is a measure of how effectively an electromagnetic wave can move in a material relative to free space, and varies with material properties. Thus our calculations put constraints on the composition of the radar-transmitting layer. The dielectric constant of pure water ice is ~3.15 [*Matsuoka et al.*, 1997] while that of vacuum (and the Martian atmosphere) is 1. *Rust et al.* [1999] found an $\varepsilon_r$ of 7.54 for pure basalt (no porosity included), while *Campbell and Ulrichs* [1969] found values up to ~9 for bulk measurements of dense basalts; for this work, we follow *Moore and Jakosky* [1989] in assuming an $\varepsilon_r$ of 8 for any lithic material. We compare our $\varepsilon_r$ estimates near the terraced craters to the modeled dielectric behavior for three-component mixtures of varying volumetric fractions of ice, basaltic rock, and air (i.e. empty pore space) using the power-law relation $\varepsilon_{mix}^{1/\gamma} = v_{rock}\varepsilon_{rock}^{1/\gamma} + v_{ice}\varepsilon_{ice}^{1/\gamma} + v_{air}\varepsilon_{air}^{1/\gamma}$ with $\gamma=2.7$, the exponent *Stillman et al.* [2010] found to best fit sand-ice mixtures (Figure 3). *Shabtaie and Bentley* [1994] showed that $\gamma=2.7$ also works for ice-air mixtures, and this fit should hold for inclusions of any dielectrically inert material (which covers the range of expected Martian materials) [*Stillman and Olhoeft*, 2008].

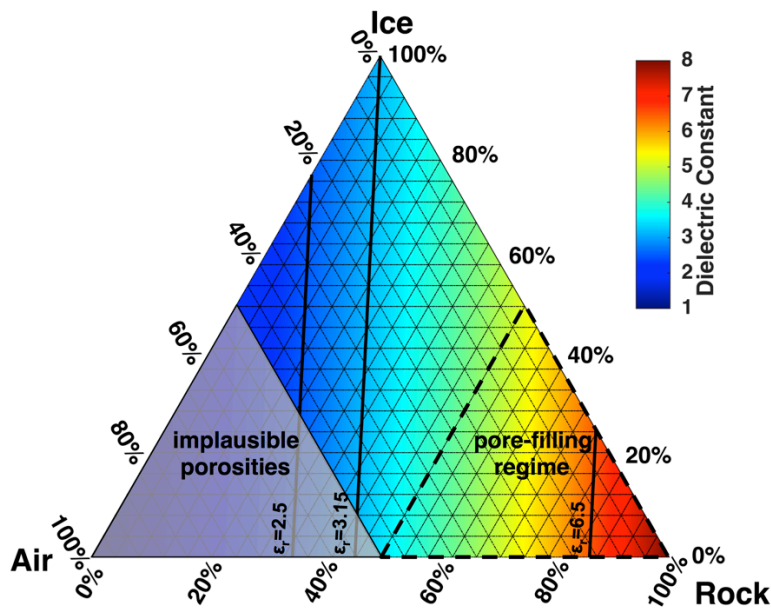

**Figure 3. Ternary diagram of dielectric constant calculations for mixtures of ice, rock and air. The shaded bottom-left corner represents the region of parameter space where porosities are implausible (>50%). The bottom-right corner inside the dashed lines is the pore-filling ice regime. Near-vertical bold lines represent contours for three dielectric constant values (2.5, 3.15 and 6.5).**

The compositions predicted by the dielectric mixing models yield non-unique solutions. For example, pure water ice has a dielectric constant of ~3.15; however, Figure 3 demonstrates that this value would also be consistent with a mixture of about equal parts rock and air or a three-component mixture of about 50% ice, 25% rock and 25% air (as well as a variety of mixtures in between).

Porosities of more than 50% (bottom-left of Figure 3) are rare in geologic materials. Porosity is expected to be at a maximum at the surface, and modeling by *Zent* *et al.* [2010] for the Phoenix landing site and measurements taken at the Viking landing sites of loose winddrift material [*Moore et al.*, 1987] suggest a surface porosity of ~50%. Windblown material is likely not representative of actual regolith though. Recent gravity measurements of porosity in the near-surface lunar regolith are significantly lower, ~10-20% [*Wieczorek et al.*, 2013; *Han et al.*, 2014]. Therefore, we adopt the Viking surface measurement as a conservative upper limit for the porosities averaged over decameters into the subsurface.

The bottom-right corner of Figure 3 represents the regime of porous regolith with varying amounts of ice filled into the pore spaces. The lines along values of constant rock content represent a continuum ranging from empty pore spaces at the bottom of the plot (the line defining 0% ice) to completely ice-filled pores (where the rock-content line intersects with 0% air).

## 3. Results

Many of the craters included in this study exhibit two terraces, with a shallower, subtler terrace within the crater wall and a wider, deeper terrace near the floor of the crater. As the shallower terraces are less pronounced, we interpret this as a subtler change in material properties – which is less likely to be detected with SHARAD. Within the area searched with CTX images, we identified 62 craters (diameter = 328 ± 129 m within one standard deviation) as single-terraced and 35 craters (diameter = 430 ± 135 m) as having two terraces. We categorized an additional 90 craters (diameter = 339 ± 271 m) as "questionably terraced".

Figure 1a shows a DTM for one such doubly-terraced crater (informally named Badger) located at 46.581°N, 194.85°E. Its large, flat terrace is at 43 m depth (which we term floor-level), and likely a major compositional interface. A smaller, shallower (17 m depth) terrace in the crater wall suggests additional complexity in the subsurface, perhaps a change in porosity or composition (such as a desiccated surface layer overlying ice). The fact that we see many craters with two terraces (which on average probe deeper than single-terraced craters) suggests that there are multiple layers within the subsurface throughout this region. The similarity in terracing across many craters indicates the same subsurface structure likely extends across the study area. However, the spread in terrace depths (Table S1) indicates that significant heterogeneity in the thicknesses of the material. Our terrace depths generally cluster into three or four depth ranges (<10 m, 17-23 m, 26-31 m and >42 m).

Our results from mapping SHARAD reflectors also show lateral heterogeneity, with greater delay times towards the south and interspersed patches that lack subsurface reflectors. We do not see any cases with two subsurface SHARAD interfaces that may be associated with the terraces, although there is occasionally a much deeper reflector beyond the depths probed by the craters (an example of which can be seen in Figure 1c). This deeper reflector is interpreted as the continuation of a buried Vastitas Borealis Formation unit often seen across Amazonis Planitia [*Campbell et al.*, 2008], which is directly south of our region of interest in Arcadia Planitia. Since this deeper reflector is related to a separate geological contact, we purposefully avoided mapping it in Arcadia Planitia, although it is possible that in rare places this reflector could have been the only one present and therefore misinterpreted. Figure 2b shows a histogram of measured one-way delay times for the entire study area, with the peak at 0.22 µs. The tail of large delay times may be contamination from this buried Vastitas Borealis Formation contact.

The extent of our mapped SHARAD reflectors is similar to that initially found by *Plaut et al.* [2009b] and *Stillman and Grimm* [2011]. We compute the area of Arcadia Planitia that contains a mapped subsurface radar reflector to be ~$2.5 \times 10^5$ km$^2$. Our map of subsurface reflectors (Figure 2a) shows a small gap near 212°E, 40°N where the surface echo is absent due to unusually high roughness (also indicted in MOLA pulse-width data) [*Carter et al.*, 2009b], preventing an accurate measure of delay time to the subsurface reflector. Similar behavior is seen by SHARAD in the Medusae Fossae Formation where the surface is extremely rough between centimeter and meter scales and the low dielectric constant limits the power of the surface return [*Carter et al.*, 2009b]. Because of this and additional reasons discussed below, the extent of the deposit shown in Figure 2a is a lower limit. We also compute a convex hull of $1.2 \times 10^6$ km$^2$ around the subsurface radar reflector measurements to put an upper limit on the area enclosed by the detections (details in Text S3).

We believe that the real geographic area covered by the deposit is closer to that calculated for the convex hull (upper bound) than for the mapped radar reflector (lower bound) because the deposit likely occurs even in reflection-free patches. We are not able to resolve any interfaces from within the vertical resolution of SHARAD (8.4 m assuming pure water ice of $\varepsilon_r=3.15$ or 5.3 m in rock with $\varepsilon_r=8$). Efforts to limit sidelobes of the surface echo can also preclude the identification of any other shallow reflectors within ~20-30 m of the surface, potentially obscuring the multiple shallower reflectors that one might otherwise expect to occur given the terracing seen within the crater walls. The deepest terraces within two craters located in areas where no SHARAD subsurface reflection appears are at 18 and 23 m depths, likely within this near-surface range where subsurface returns may be masked by the bright surface return or attenuated in the processing to limit sidelobes. Because we find terraced craters in some of these subsurface-reflector-free patches, it is probable that the layer does not disappear when the reflector disappears but rather the interface becomes too shallow for its base to be resolved by the radar. Reflection-free patches could also be produced by a small dielectric contrast between overlying and underlying material, or a gradual vertical dielectric transition.

Two SHARAD tracks pass near the crater shown in Figure 1 with one-way delay times of the subsurface reflector ranging from 0.21 µs to 0.27 µs. Using an inverse-distance weighted-mean of radar returns within 10 km gives a one-way delay time of 0.24 µs relative to the surrounding surface. If the SHARAD interface corresponds to the small, shallower terrace at 17 m depth, then $\varepsilon_r = 17.9 \pm 0.39$, an unrealistically large value for geologic materials. Thus, it is implausible that the radar reflector corresponds to this upper compositional interface. However, associating this delay time with the second, deeper terrace at a depth of 43 m yields a realistic $\varepsilon_r \sim 2.9 \pm 0.39$. This is consistent with our expectation that the deeper, more significant, terraces are more likely to correspond to the radar reflection. Therefore, we compare the deeper terraces to subsurface radar returns. Four out of the 11 terraced craters for which we have DTMs exhibit two terraces (important because we require the deeper terrace) *and* have subsurface radar reflectors within 10 km (important because of the lateral heterogeneity in the thickness of the layer). For three of these craters, we find similar dielectric constants of 2.1, 2.6, and 2.9.

The fourth crater yields a higher dielectric constant of 6.5 (right-most contour in Figure 3), but this crater is likely too shallow to provide realistic interpretations of the layer. An $\varepsilon_r$ of 6.5 is a reasonable value for pore-filling ice or porous basalt, a common material on Mars that is often layered. However, this crater's deepest terrace is at 31 m whereas the other three have floor terraces deeper than 42 m. Thus, this fourth crater likely does not excavate to sufficient depth to intersect the radar reflector interface despite it having two terraces, and the measurements from the other three craters are expected to be most representative of the bulk composition of the layer.

The average dielectric constant of the three deepest craters is 2.5 with an error of 0.28. Because our $\varepsilon_r$=2.5 contour is clearly outside the pore-filling ice parameter space (Figure 3), we conclude that the material in the upper decameters of the surface is excess ice of up to 75% volumetric fraction ice. Because we can only compute a bulk dielectric constant from surface to base of the radar-detected subsurface interface, this value likely includes the effects of a desiccated regolith layer at the surface, which could bias the interpretation towards lower ice contents. Our results can only place constraints on the overall composition, ignoring the finer structure that likely exists in the upper decameters of the subsurface. Using this average dielectric constant, we convert delay times to depths to constrain the volume of the deposit to be between ~$1.3 \times 10^4$ km$^3$ and $6.1 \times 10^4$ km$^3$ (see S3 for details).

For the $\varepsilon_r$=2.5 contour, the maximum allowed volumetric rock fraction is ~20%, meaning the initial porosity would have to be 80% if the ice were deposited via the usual pore-filling ice mechanism. Even at the high-end of our error bars, where the maximum rock fraction reaches 30%, the initial porosity would need to be ~70%. These are unreasonable initial porosities, and thus we rule out the atmospheric vapor pore-filling ice origin.

Dielectric constants as low as ~3.0 can be due to low-density volcanic materials such as pumice, tuff and ash [*Campbell and Ulrichs*, 1969]. Such dielectric constants have been measured in the upper few hundred meters of the Medusae Fossae Formation, and were interpreted to be a low-density pyroclastic unit [*Watters et al.*, 2007; *Carter et al.*, 2009b]. While similarly porous volcanic materials could explain the dielectric constants in Arcadia Planitia, one cannot explain other morphological features such as expanded craters [*Viola et al.*, 2015] or the ice-exposing craters [*Dundas et al.*, 2014] with such materials.

Carbon dioxide ice has a dielectric constant of ~2.1 [*Pettinelli et al.*, 2003]. While our calculations yield numbers that could also be consistent with $CO_2$ ice and lithic mixtures, the temperatures decameters into the subsurface will equal the annual mean surface temperature, which is much greater than the frost point for $CO_2$ ice. Thus, conditions at the mid-latitudes are not stable for perennial buried $CO_2$ ice and we rule out this composition.

The SHARAD subsurface reflectors we detect across Arcadia Planitia are of similar brightness to the surface reflector, suggesting this subsurface layer does not strongly attenuate the signal, behaving similarly to reflectors in the polar layered deposits [*Putzig et al.*, 2009] and the base of lobate debris aprons [*Holt et al.*, 2008; *Plaut et al.*, 2009a]. This provides additional support in favor of an ice-rich layer rather than a lithic-rich layer, as lithic materials cause more attenuation of the radar return [*Carter et al.*, 2009a, 2009b]. Although there are numerous interpretations that would match our calculations, the prevalence of surface features associated with the presence of excess ice across Arcadia Planitia [*Holt et al.*, 2008; *Byrne et al.*, 2009; *Kadish et al.*, 2009; *Plaut et al.*, 2009a; *Dundas et al.*, 2014; *Viola et al.*, 2015] also supports our ice interpretation.

**4. Conclusions and Discussion**

We present evidence of widespread subsurface layering from SHARAD radar interfaces and terraced craters across Arcadia Planitia. We use HiRISE-derived digital terrain models of terraced craters to directly measure the depths to this SHARAD interface. From the combination of depths and radar delay times, we calculate dielectric constants, and infer that the decameters of material between the surface and interface is mostly water ice. Our work constrains depths to the presumed base of the ice layer (mode depth = 42 m, mean depth = 51 m, standard deviation = 18 m, using our derived dielectric constant of 2.5). We also place constraints on the area covered by the ice ($<1.2 \times 10^6$ km$^2$) over a latitude range of 38°N - 52°N, and place limits on the volume of ice ($1.3 \times 10^4$ km$^3$ - $6.1 \times 10^4$ km$^3$). Superposed expanded secondary craters in the region also support the presence of excess ice, and dating of their primary craters suggests that this ice has been present at this location for 10s of Myr [*Viola et al.*, 2015].

Climate models suggest that widespread deposition of ice could have occurred in the mid-latitudes during Mars' high obliquity (>35°) periods under certain conditions; e.g. *Madeleine et al.* [2009] found ~10 mm/year of ice accumulation was possible in Arcadia Planitia when equatorial "tropical mountain glacier" ice is available to be mobilized, atmospheric dust opacity is high ($\tau=2.5$), and obliquity is 35°. Any such period of ice deposition/snowfall would likely need to be quickly followed by the formation of a protective lag to preserve the layer. However, uncertainty exists in conditions needed to preserve the layer for the 10s of Myr suggested by *Viola et al.* [2015].

The response of ice to changing conditions is dependent on the diffusion coefficient of water vapor through the regolith, with higher diffusion coefficients corresponding to faster responses. Experimental studies of ice diffusion properties under simulated Martian conditions suggest Martian regolith has high diffusion coefficients [*Chevrier et al.*, 2007; *Hudson et al.*, 2007; *Bryson et al.*, 2008; *Sizemore and Mellon*, 2008]; in other words, water vapor should readily diffuse from the subsurface during even brief excursions through periods of ice instability. Evidence that ice in the upper meter of the surface responds rapidly to climatic changes comes from Mars Odyssey observations [*Boynton et al.*, 2002; *Feldman et al.*, 2002; *Mitrofanov et al.*, 2002], which show a match between today's ice distribution and current climate model predictions [*Mellon et al.*, 2004; *Schorghofer and Aharonson*, 2005]. Depths to ice at the Phoenix site suggest the extent of ground ice is in approximate equilibrium with the present climate [*Mellon et al.*, 2009].

With such findings, a mechanism to preserve an ice sheet for 10s of Myr remains elusive. Experiments by *Hudson et al.* [2007] have difficulty attaining the low diffusion coefficients needed to preserve ice for this long. Laboratory experiments on mass loss rates of pure water ice by *Bryson et al.* [2008] find that a 1 m layer of ice underneath a 2 m layer of fine-grained basaltic regolith could remain stable for 400 Kyr (since the last large obliquity change). It is unclear from their study, however, if thicker ice could remain through the hundreds of predicted large obliquity changes over the course of last 10s of Myr. *Head et al.* [2003] and *Schorghofer and Forget* [2012] predict that a thick ice sheet would be geologically recent and actively retreating to come into equilibrium with the current atmospheric conditions at these mid-latitudes. For example, ice stability models by *Schorghofer and Forget* [2012] found that a 4.5 Myr old ice sheet (an order of magnitude younger than this ice) 30 m thick and extending to the equator would not have survived at these latitudes to the present day.

While atmospheric water vapor can recharge the ice within pore spaces quickly, once excess ice has sublimated away, it cannot be recharged via atmospherically-derived water. *Fisher* [2005] proposed a mechanism by which thermal contraction of pore-ice works to open voids that can be further filled with ice through diffusion of water vapor. His results suggest that the upper few meters of Martian regolith could reach up to 70% ice by mass in 10 Myr. But it is doubtful this mechanism could form the thicknesses of excess ice at the depths we infer to be present across Arcadia Planitia.

Segregation of pre-existing pore ice, analogous to frost heave, has also been proposed for generating excess ice near the surface by *Sizemore et al.* [2015], who modeled ice lens initiation and development, tracking phase partitioning of water in soil pores. Their simulations suggest ice lens formation is possible in the upper tens of centimeters of the surface in fine-grained soils. However, the thicknesses of ice discussed in this paper are likely too great to form from ice lenses, even under idealized soil and temperature conditions with continuous growth of segregated ice.

We consider snowfall to be the leading formation mechanism proposed for this ice deposit because it can most easily explain the thickness and widespread nature of the excess ice observed. An ice deposition episode similar to that proposed by *Madeleine et al.* [2009] would be a good candidate for the excess ice across Arcadia Planitia, but because of the expected age of the ice, there is a need for a theoretical paradigm to explain its preservation. The latitude range of this ice deposit, its purity, thickness, area and volume provide important constraints for future climate and ice stability models over the last 10s of Myr.

**Acknowledgements**
SHARAD radar tracks were processed using the Colorado SHARAD Processing System (CO-SHARPS) and analyzed using geophysical interpretation software provided courtesy of SeisWare International Inc. Access to the CO-SHARPS processing boutique can be requested at https://www.boulder.swri.edu/sharad.php. All HiRISE and CTX images used are publically available through NASA's Planetary Data System (https://pds.nasa.gov/). Eight of the Digital Terrain Models AMB created for this project have been archived and released to the public (http://www.uahirise.org/dtm/), and the other three DTMs will be released by January 2016. We benefitted from insightful conversations from Hanna Sizemore, David Stillman and James T. Keane. AMB was supported by the NSF Graduate Research Fellowship under Grant No. DGE-1143953.

---

Supporting Information for

**Widespread Excess Ice in Arcadia Planitia, Mars**

Ali M. Bramson[1], Shane Byrne[1], Nathaniel E. Putzig[2], Sarah Sutton[1], Jeffrey J. Plaut[3], T. Charles Brothers[4] and John W. Holt[4]

[1]Lunar and Planetary Laboratory, University of Arizona, Tucson, Arizona, USA
[2]Southwest Research Institute, Boulder, Colorado, USA
[3]Jet Propulsion Laboratory, Pasadena, California, USA
[4]Institute for Geophysics, University of Texas at Austin, Austin, Texas, USA

**Contents of Supporting Information**

Text S1 to S3
Figures S1 to S3
Table S1

**Introduction**

This file contains information on the exclusion of radar sidelobes and clutter in our analysis of the SHARAD subsurface interfaces as well as details on how we estimate the error in our dielectric constant calculations. Also included is a figure showing all 11 craters for which we made Digital Terrain Models (DTMs) and a table of the measurements and calculations involved with each crater and DTM, including latitude, longitude, diameter, and HiRISE stereo pair image IDs used for each crater; elevations, depths, dips and elevation range for the surroundings, terrace(s), and bottom of the crater's nested pit (if applicable); number of, mean, weighted mean, and standard deviation of SHARAD measurements; dielectric constant calculations (based on terrace depths and weighted mean of SHARAD subsurface points within 10 km) and errors.

**Text S1. SHARAD Processing, Clutter Simulations and Sidelobe Avoidance**

Colorado SHARAD Processing System (CO-SHARPS) parameters used in the processing of SHARAD radargrams: an aperture length of 2048 (12-second aperture), Hann weighting, a peak SNR of 32 dB, and a 0.2 Hz bandwidth window.

Radar "clutter" consists of bright reflections from off-nadir surface topography, giving late returns that may be mistaken for subsurface returns. In our mapping, we compared radargrams to their corresponding clutter simulations produced using Mars Orbiter Laser Altimeter (MOLA) data, the radar trajectory and the effects of radar focusing [*Holt et al.*, 2006, 2008]. When a subsurface interface appeared in the clutter simulation, we did not map it as a subsurface interface (Figure S1).

Because of the band-limited nature of the radar signal, processing produces sidelobes adjacent in delay time to the local maxima in the return power of the SHARAD signal, and the trailing sidelobes of strong surface reflections can be mistaken for shallow subsurface reflectors [*Seu et al.*, 2007]. In processing, SHARAD chirp compression places the first sidelobe at a one-way delay of 0.12 μs relative to the surface reflector (and weaker by 20 dB) and a second sidelobe at a relative one-way delay of 0.21 μs (and 34 dB weaker). We used a Hann filter, which suppresses sidelobes, in the processing of the radargrams; however, if we found that a late reflector mirrored the surface reflector exactly and occurred near these delay times, we assumed it was due to the instrument's sidelobes and we did not mark it as a subsurface return.

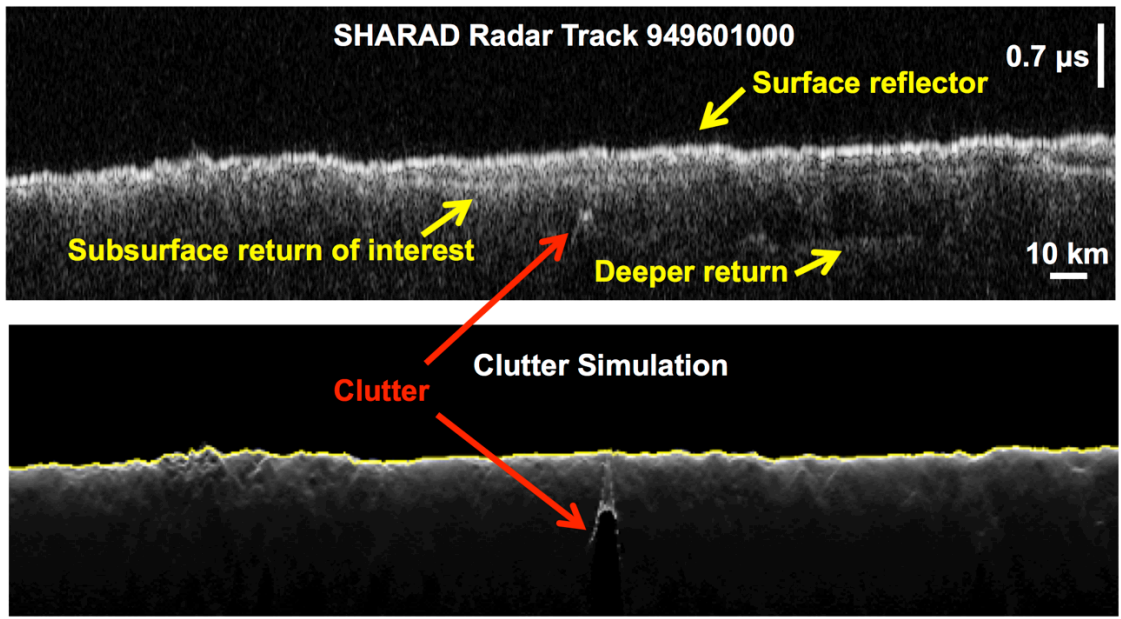

**Figure S1.** Radargram (top) and corresponding clutter simulation (bottom). Yellow arrows point to the surface and subsurface interfaces, which have no corresponding signature in the clutter simulation. Red arrows point to "clutter" which appear in the clutter simulation and the radargram. Clutter was not mapped, as it is caused by off-nadir topography rather than subsurface interfaces.

**Text S2. Estimation of Errors in Dielectric Constant**

*Christian et al.* [2013] suggested that errors in radar-derived layer thickness are dominated by the SHARAD time-sampling resolution of 0.0375 μs. Propagating this through to an error in the dielectric constant would give an error in $\varepsilon_r$ of ~0.31. We take a more conservative approach by using the full width at half maximum (FWHM) of the power of the surface and subsurface reflectors in a profile averaged over 50 pixels taken at the nearest point to the crater in the track that passes closest to the crater (Figure S2). Power values used are DN values from radargram images, which were scaled from SNR values between -3 dB and 32 dB. We use the FWHM values to calculate one-sigma uncertainties of each reflector (surface and subsurface) and propagate the combined uncertainty of the time difference through to the dielectric constant measurements using the mean delay times and the deeper terrace depth for that crater.

The vertical precision in HiRISE DTMs is generally better than 1 m [*Sutton et al.*, 2015] and so, by comparison to the radar delay-time error, is insignificant, though it is included in our error calculations. Our calculations of full width at half maximum (shown in Figure S2) give a one sigma uncertainty of delay-time difference between surface and subsurface reflectors of 0.047-0.066 μs which dominate the error in our dielectric constants, and leads to an uncertainty therein of 0.39-0.61 in $\varepsilon_r$.

The average dielectric constant for craters 0, 2 and 7 (those that are deep enough to likely go through the entire layer) give an average dielectric constant of 2.5. The error in this average is reduced to $\sqrt{(0.39^2 + 0.61^2 + 0.43^2)} / 3$ to give our resulting
$\varepsilon_r = 2.5 \pm 0.28$.

Details of each error calculation:

Crater 0: The power is taken as an average of the columns of radargram 949601000 centered on pixel 3699 ± 25 pixels.
    Error in DTMs < 1 m è $\sigma_{\Delta x}$ = 1.4 m
    Error in SHARAD:
        Surface σ = 0.037 μs
        Subsurface σ = 0.029 μs
        $\sigma_{\Delta t}$ = 0.047 μs
    Propagate Errors to Dielectric Constant
        $\sigma_{\varepsilon r \; crater \; 0}$ = 0.39

Crater 2: The power is taken as an average of the columns of radargram 676701000 centered on pixel 1512 ± 25 pixels.
    Error in DTMs < 1 m è $\sigma_{\Delta x}$ = 1.4 m
    Error in SHARAD for Crater 2:
        Surface σ = 0.061 μs
        Subsurface σ = 0.020 μs
        $\sigma_{\Delta t}$ = 0.064 μs
    Propagate Errors to Dielectric Constant
        $\sigma_{\varepsilon r \; crater \; 2}$ = 0.61

Crater 7: The power is taken as an average of the columns of radargram 507901000 centered on pixel 209 ± 25 pixels.
    Error in DTMs < 1 m è $\sigma_{\Delta x}$ = 1.4 m
    Error in SHARAD for Crater 2:
- Surface σ = 0.062 μs
- Subsurface σ = 0.022 μs
- $\sigma_{\Delta t}$ = 0.066 μs

    Propagate Errors to Dielectric Constant

- σ_εr crater 7 = 0.43

Crater 10: The power is taken as an average of the columns of radargram 577801000 centered on pixel 65 ± 25 pixels.
    Error in DTMs < 1 m è σ_Δx = 1.4 m
    Error in SHARAD for Crater 2:
        Surface σ = 0.053 μs
        Subsurface σ = 0.037 μs
        σ_Δt = 0.064 μs
    Propagate Errors to Dielectric Constant
- σ_εr crater 10 = 0.49

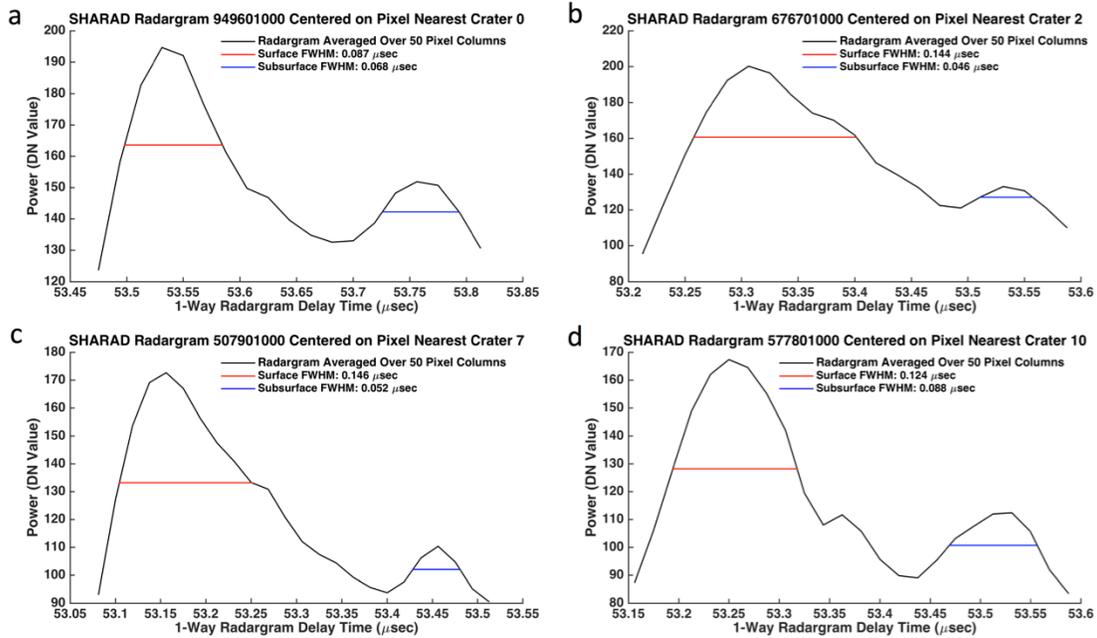

**Figure S2.** Average power plotted with one-way delay times for each of the radar tracks nearest to the four craters. We calculated full width at half maximums of the surface (red) and subsurface (blue) returns to find the one-sigma uncertainties in delay time (and thus dielectric constant) calculations.

**Text S3. Area and Volume Calculations**

The area of Arcadia Planitia that contains a mapped subsurface radar reflector is ~2.5 x $10^5$ km$^2$ if we bin the data onto a grid of ~10 km x ~10 km cells and sum the area of the grid containing a subsurface reflector. A robust lower limit on the ice volume of 1.3 x $10^4$ km$^3$ comes from multiplying the mean measured delay time for each cell, converted to a depth using $\varepsilon_r$ = 2.5, by the area of that cell for each grid point that contains at least one subsurface radar measurement. This method ignores the areas between SHARAD detections that likely contain ice too thin to be detected by SHARAD (< ~20 – 30 m).

We place an upper limit on the volume of ice by multiplying the mean delay-time (0.27 μs) converted to depth assuming $\varepsilon_r$ = 2.5 (51 m) by the convex hull calculated around the subsurface radar measurements within Arcadia Planitia of 1.2 x $10^6$ km$^2$. Using the mean depth from SHARAD measurements overestimates the actual mean depth because there are shallow interfaces at sufficiently small delay times that we can't measure. The convex hull area is also an upper limit on the area of the deposits as it assumes all the interior gaps are filled with the layer. This calculation yields an upper limit on the volume of ice of 6.1 x $10^4$ km$^3$, which would correspond to a volume of proportions similar to a ~360 m layer of ice covering an area the size of Wisconsin.

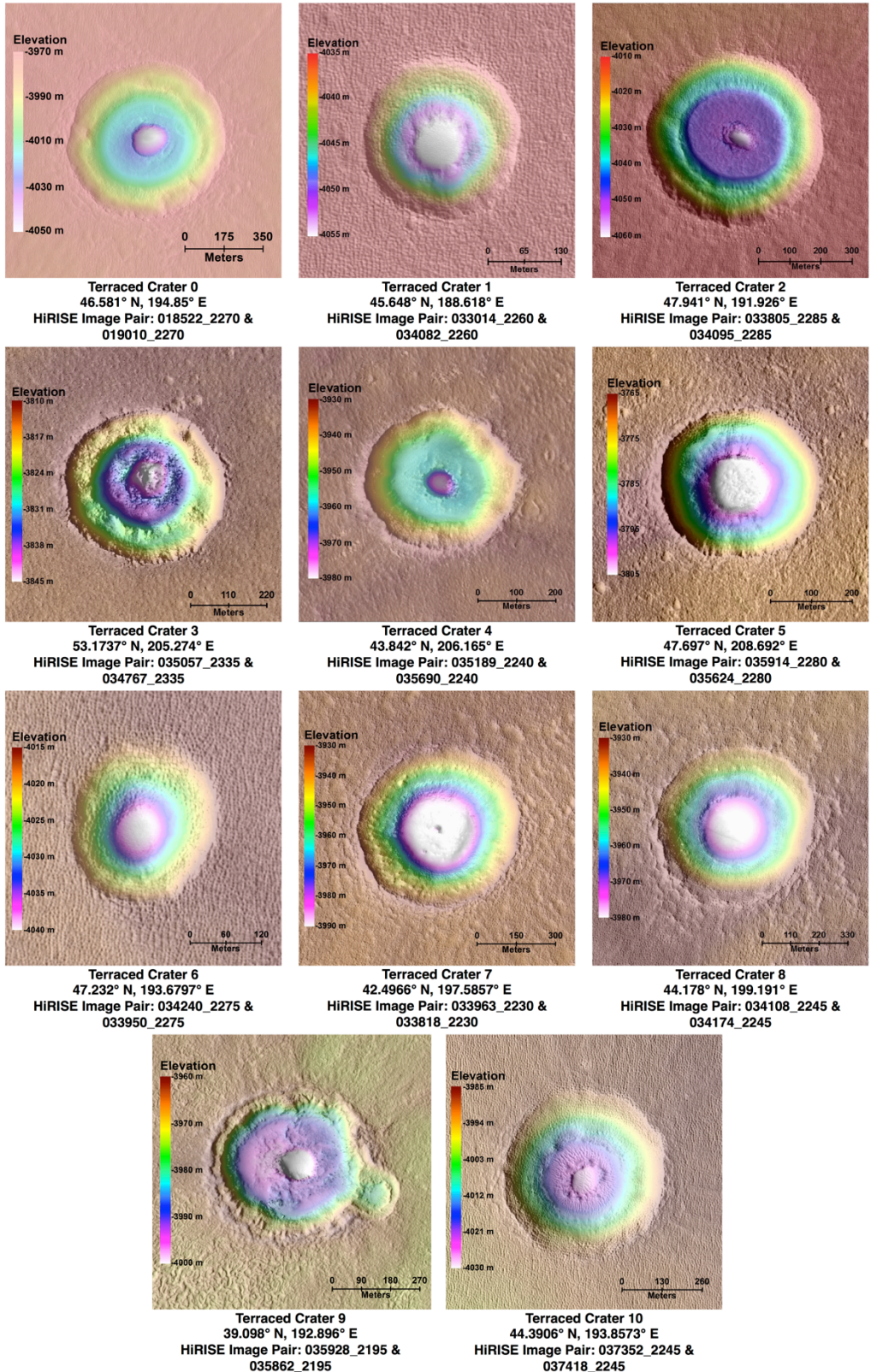

**Figure S3.** Digital terrain models (with shaded relief) made for the study along with the crater ID given for the study, its location and the image pair used to make each DTM. Colors represent the elevations across the DTM.

| | Terraced Crater | | | | | | | | | | |
|---|---|---|---|---|---|---|---|---|---|---|---|
| Terraced Crater Number Label | 0 | 1 | 2 | 3 | 4 | 5 | 6 | 7 | 8 | 9 | 10 |
| HiRISE Image Pair for DTM | 018522_2270 019010_2270 | 033014_2260 034082_2260 | 033805_2285 034095_2285 | 035057_2335 034767_2335 | 035189_2240 035690_2240 | 035914_2280 035624_2280 | 034240_2275 033950_2275 | 033963_2230 033818_2230 | 034108_2245 034174_2245 | 035928_2195 035862_2195 | 037352_2245 037418_2245 |
| Diameter (m) | 734 | 292 | 579 | 461 | 391.127 | 394 | 258 | 631 | 625 | 504 | 535 |
| Latitude (°N) | 46.58 | 45.65 | 47.94 | 53.17 | 43.84 | 47.70 | 47.23 | 42.50 | 44.18 | 39.10 | 44.38 |
| Longitude (°E) | 194.85 | 188.62 | 191.93 | 205.27 | 206.17 | 208.69 | 193.68 | 197.59 | 199.19 | 192.90 | 193.86 |
| Surroundings Elevation (m) | -3975.5 | -4033.0 | -4006.3 | -3812.2 | -3931.0 | -3768.9 | -4015.3 | -3934.7 | -3934.3 | -3968.5 | -3990.6 |
| Shallow Terrace Elevation (m) | -3992.7 | -4038.8 | -4035.6 | -3821.2 | -3957.5 | -3795.2 | -4022.6 | -3943.6 | -3960.2 | -3965.6 | -4008.5 |
| Deeper Terrace Elevation (m) | -4018.3 | -4050.6 | -4048.1 | -3835.5 | | | -4034.1 | -3991.1 | | -3991.4 | -4021.3 |
| Bottom Pit Elevation (m) | -4051.2 | -4058.4 | -4056.6 | -3843.7 | -3977.4 | -3808.0 | -4041.7 | no pit | -3984.2 | -4010.6 | -4027.5 |
| Shallow Terrace Depth (m) | 17.2 | 5.8 | 29.3 | 9.0 | 26.4 | 26.3 | 7.3 | 8.8 | 25.9 | 2.8 above | 17.9 |
| Deeper Terrace Depth (m) | 42.8 | 17.6 | 41.8 | 23.3 | | | 18.8 | 56.3 | | 23.0 | 30.7 |
| Bottom Pit Depth (m) | 75.7 | 25.4 | 50.3 | 31.5 | 46.3 | 39.1 | 26.4 | | 49.9 | 42.1 | 36.9 |
| Surroundings Dip | 0.12° | 0.05° | 0.68° | 0.13° | 0.06° | 0.09° | 0.34° | 0.16° | 0.34° | 0.13° | 0.31° |
| Shallow Terrace Dip | 0.19° | 1.11° | 0.61° | 1.58° | 0.26° | 0.89° | 0.68° | 0.41° | 1.82° | 0.85° | 2.35° |
| Deeper Terrace Dip | 0.75° | 1.42° | 0.19° | 0.57° | | | 5.83° | 0.85° | | 1.22° | 0.76° |
| Surroundings Elevation Range (m) | 11.6 | 3.7 | 22.0 | 7.6 | 5.2 | 7.7 | 6.7 | 9.7 | 17.1 | 11.8 | 12.7 |
| Shallow Terrace Elevation Range (m) | 10.6 | 6.6 | 8.2 | 15.4 | 3.6 | 7.3 | 4.3 | 9.5 | 13.3 | 7.1 | 13.5 |
| Deeper Terrace Elevation Range (m) | 8.8 | 4.4 | 3.0 | 7.0 | | | 5.1 | 6.4 | | 9.7 | 4.9 |
| Bottom Pit Elevation Range (m) | 7.7 | 0.4 | 2.4 | 4.0 | 1.7 | 0.7 | 3.3 | | 1.4 | 2.9 | 1.1 |
| # SHARAD Track Points Measured within 10 km | 25 | 129 | 69 | 130 | 143 | 57 | 135 | 133 | 210 | 53 | 84 |
| # Points with Subsurface Interface within 10 km | 25 | 0 | 33 | 0 | 13 | 37 | 0 | 93 | 111 | 0 | 61 |
| Mean of Subsurface Delays within 10 km (µs) | 0.24 | | 0.21 | | | | | 0.31 | | | 0.27 |
| Standard Deviation of Subsurface Delays within 10 km (µs) | 0.02 | | 0.04 | | | | | 0.02 | | | 0.03 |
| Weighted Mean of Subsurface Delays within 10 km (µs) | 0.24 | | 0.20 | | | | | 0.30 | | | 0.26 |
| Dielectric Constant using Weighted Mean within 10 km | 2.94 | | 2.06 | | | | | 2.62 | | | 6.46 |
| Error in Dielectric Constant | 0.39 | | 0.61 | | | | | 0.43 | | | 0.49 |

**Table S1.** Table of all results, including properties of each terraced crater for which we generated DTMs and the corresponding SHARAD one-way delay time measurements for the craters that had subsurface radar interfaces within 10 km. Also listed are the dielectric constant calculations based on an inverse-distance weighted mean of one-way delay time measurements within 10 km for craters that had a second terrace. Blue columns represent craters that have two terraces *and* subsurface radar interfaces within 10 km of the crater.

**Supporting Information References**